\documentclass[prl,showpacs,twocolumn,amsmath,amssymb]{revtex4}

\usepackage{graphicx}

\newcommand{\dt}{\Delta\tau}
\newcommand{\vk}{\mathbf{k}}

\begin{document}

% based on v4, mostly reverted to v3 (no DDMRG comparison)

\title{Numerically exact Green functions from Hirsch-Fye quantum Monte Carlo simulations}

\author{N.~Bl\"umer}
\email{Nils.Bluemer@uni-mainz.de}
\affiliation{Institute of Physics, Johannes Gutenberg University, 55099 Mainz, Germany}

\date{\today}

  \begin{abstract}
    We present a new method for extracting numerically exact
    imaginary-time Green functions from standard Hirsch-Fye quantum
    Monte Carlo (HF-QMC) simulations within dynamical mean-field
    theory. By analytic continuation, angular resolved spectra are
    obtained without the discretization bias previously associated
    with HF-QMC results. The method is shown to be accurate even at
    very low temperatures ($T=W/800$ for bandwidth $W$) in the
    strongly correlated regime.
  \end{abstract}
  \maketitle

%%%%%%%%%%%%%%%%%%%%%%%%%%%%%%%%%%%%%%%%%%%%%%%%%%%%%%%%%%%%%%%%%%%%%%%%
% Introduction
%%%%%%%%%%%%%%%%%%%%%%%%%%%%%%%%%%%%%%%%%%%%%%%%%%%%%%%%%%%%%%%%%%%%%%%%

  Photoemission spectroscopy (PES) and related techniques (inverse
  PES, X-ray absorption spectroscopy) are among the most useful
  experimental probes of electronic properties of solids \cite{PES}.
  Already the angle- and spin-integrated variants can characterize
  materials as weakly or strongly correlated metals or as insulators.
  In addition, angle-resolved photoemission spectroscopy (ARPES) gives
  a nearly direct view on the occupied part of the electronic band
  structure. However, the interpretation of such experimental data and
  the separation of surface from bulk contributions is by no means
  trivial. Thus, both a reliable analysis of the experiments and an
  understanding of the underlying physics require comparisons with
  theoretical predictions. Often, good agreement is found with spectra
  obtained within density functional theory (DFT), e.g.,
  within the local density approximation (LDA) \cite{DFT}. For strongly
  correlated materials, however, the LDA and/or the interpretation of
  the DFT parameters as many-body dispersion break down; then, true
  many-body approaches, usually based on Hubbard-type models, are required.
  
  A nonperturbative treatment of Hubbard-type models for correlated
  electron systems is possible by iterative solution of the DMFT
  self-consistency equations \cite{Georges96} using quantum Monte
  Carlo (QMC) methods. In the Hirsch-Fye QMC algorithm
  \cite{Hirsch86}, the imaginary-time path integral is discretized
  into $\Lambda$ time slices of uniform width $\dt=\beta/\Lambda$; a
  Hubbard-Stratonovich (HS) transformation replaces the
  electron-electron interaction at each time step by a binary
  auxiliary field which is sampled by standard Markov Monte Carlo
  techniques. After convergency, estimates of the local imaginary-time
  Green function $G(\tau)$ are obtained (only) on the grid
  $\tau_l=l\dt$ with ($0\le l\le\Lambda$).  For the computation of
  spectral functions, this data has to be continued to the real axis,
  typically using maximum entropy methods (MEM) \cite{Jarrell96};
  these regularize the ill-conditioned inversion of the equation
  \begin{equation}\label{eq:Gtau}
    G(\tau)=\int_{-\infty}^{\infty}d\omega \,
    \frac{\exp(-\tau\omega)}{1+\exp(-\beta\omega)} A(\omega)
  \end{equation}
  by finding a spectrum $A(\omega)$ that is as smooth (or similar to a
  chosen default model) as possible within the constraints set by the
  data $\{G(\tau_l)\}$ and its error bars. If needed, the real part of
  the local Green function is then obtained by Kramers-Kronig
  transformation of Im$G(\omega)=-\pi A(\omega)$. Finally, momentum
  resolved spectra corresponding to ARPES measurements may be obtained
  via the real-frequency self-energy $\Sigma(\omega)$.

  While impressive results have been obtained using the QMC/MEM
  procedure as outlined above, e.g., in the context of ``kink''
  anomalies in ARPES spectra \cite{Byczuk07}, it has one fundamental
  problem: the Hirsch-Fye QMC estimates of the imaginary-time Green
  function contain systematic Trotter errors which are typically much
  larger than the statistical errors, often by orders of magnitude.
  However, the MEM takes only the statistical errors into account,
  partially by quite elaborate formalism, while the larger systematic
  errors in the data are neglected.  Consequently, all resulting
  spectra are biased, to an unknown extent, by the Trotter
  discretization.

  In this Letter, a method is proposed which essentially eliminates
  these problems: using a novel extrapolation scheme, we extract
  continuous estimates of the imaginary-time Green function without
  significant Trotter error from conventional HF-QMC data; these are
  then proper bases for analytic continuation techniques. The method
  works well even at very low temperatures and for comparatively
  coarse imaginary-time discretizations. These properties make it very
  attractive for future calculations of spectra, e.g., in {\em ab
    initio} LDA+DMFT studies; they also establish that -- contrary
  to common belief -- reliable HF-QMC results do not depend on a good
  resolution of the rapid initial decay of $G(\tau)$.

%%%%%%%%%%%%%%%%%%%%%%%%%%%%%%%%%%%%%%%%%%%%%%%%%%%%%%%%%%%%%%%%%%%%%%%%
% Method and results for T=1/200
%%%%%%%%%%%%%%%%%%%%%%%%%%%%%%%%%%%%%%%%%%%%%%%%%%%%%%%%%%%%%%%%%%%%%%%%

  {\it Raw HF-QMC results --} In the following, the method will be
  defined and illustrated using a quite ambitious example: the
  half-filled single-band Hubbard model (semi-elliptic density of
  states with bandwidth $W=4$) at the very low temperature $T=1/200$
  for $U=4.95$, i.e., in the strongly correlated metallic regime. As
  HF-QMC results had so far only been reported for temperatures
  $T\gtrsim 1/50$, these parameters have been believed to be out of
  reach of the Hirsch-Fye QMC method \cite{Werner06}.

\begin{figure}
    \includegraphics[width=\columnwidth]{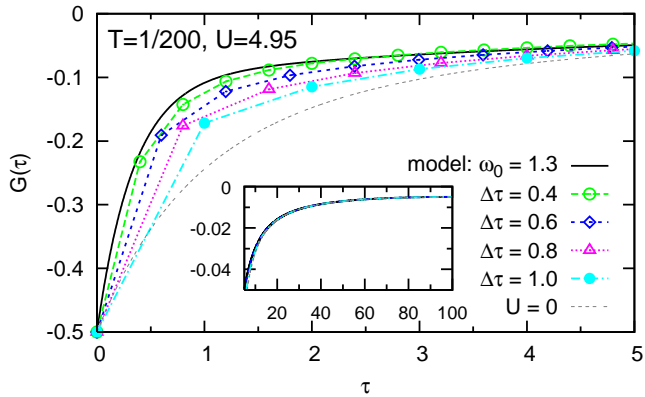}
    \caption{(Color online) Imaginary-time Green function for
      $T=1/200$, $U=4.95$: HF-QMC estimates for different
      discretizations $\dt$ (symbols); lines are guides to the eye
      only. Also shown: Green functions corresponding to a model
      self-energy (solid line) and to the noninteracting limit (thin
      dashed line); inset: same data up to symmetry
      point $\tau=\beta/2$.}\label{fig:GtauB200}
  \end{figure}

  Discrete HF-QMC estimates of the imaginary-time
  Green function are shown as symbols in the main panel of Fig.\
  \ref{fig:GtauB200}
  for relatively large values of the discretization $\dt\in [0.4,1]$.
  Clearly, the Trotter errors in these data sets are very significant
  at least for $\tau\lesssim 5$, much larger than the statistical
  and convergency errors (of about $10^{-4}$). However, extrapolation
  schemes that have successfully been used for observables such as
  quasiparticle weight $Z$, energies, and double occupancies
  \cite{Bluemeretc,Bluemer07} cannot directly be applied here, since
  the $\tau$ grid is different for each data set. In addition, none of
  the data sets really covers the initial rapid drop of $|G(\tau)|$:
  even for the finest discretization $\dt=0.4$, $|G(\tau)|$ is already
  halved at the first nontrivial data point. In fact, according to the
  assumption \cite{Werner06} that a good resolution of the highly
  curved initial region was necessary, HF-QMC results for such coarse
  grids should not yield useful information at all; instead one would
  have to resort to much finer grids of $\dt\lesssim 1/(5U)\approx
  0.04$ \cite{Werner06}.

  These considerations, however, overlook an important physical point:
  The behavior of $G(\tau)$ at small $\tau$ is restricted by
  second-order weak-coupling perturbation theory (which is
  exact for the curvature of $G$ at $\tau\to 0$ in the half-filled case; see below and Fig.\
  \ref{fig:GderivB200}). Consequently, reasonable accuracy at small
  $\tau$ can be expected from a model Green function based on a (here
  particle-hole symmetric) two-pole approximation to the self-energy,
  $\Sigma_{\text{model}}(\omega)= -A\omega/(\omega^2-\omega_0^2)$;
  since the weight $A=U^2/4$ is fixed by the leading large-frequency
  asymptotics \cite{Potthoff97}, only the position of the poles (at
  $\pm \omega_0$) remains as a free parameter. As indicated by the
  solid line in Fig.\ \ref{fig:GtauB200}, this simple model (for
  $\omega_0=1.3$) has all the features that we expect from the true
  Green function, including the rapid initial drop; in fact, a visual
  inspection of the trends suggests that it better approximates the
  true Green function (for small $\tau$) than any of the QMC data
  sets.  Still, as we will show in the following, the latter contain
  the necessary information for computing very precise estimates of
  the true Green function at all $\tau\in[0,\beta]$; the model will
  only be needed for interpolating the raw QMC data.

\begin{figure}
    \includegraphics[width=\columnwidth]{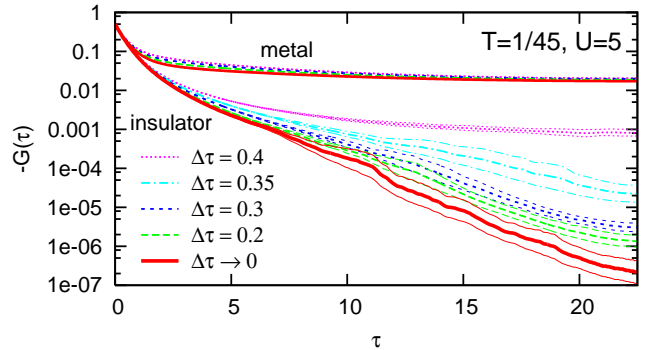}
    \caption{(Color online) HF-QMC estimates of imaginary-time Green
      functions for $T=1/45$, $U=5$ for the metallic (upper set of
      curves) and insulating (lower set) phases; side bands (thinner lines)
  indicate corresponding error estimates. }\label{fig:GtauB45U5}
  \end{figure}

  {\it Extrapolation procedure --} As a starting point for the
  extrapolation, we need accurate QMC data $\{G(\tau_l)\}$ with
  reliable error bars.  While {\it arithmetic} averages are clearly
  appropriate for combining the Green functions obtained in parallel
  QMC runs for the {\it same} impurity model, errors can be minimized
  (most notably in insulating phases) by using {\it geometric}
  averages for combining estimates of $\{G(\tau_l)\}$ from {\it
    different DMFT iterations}. In practice, we transform all data to
  $\log[-G(\tau)]$ before taking arithmetic averages; the appropriateness of
  this logarithmic scale will become apparent below (cf.\ Fig.\
  \ref{fig:GtauB45U5}). 
  
\begin{figure}
    \includegraphics[width=\columnwidth]{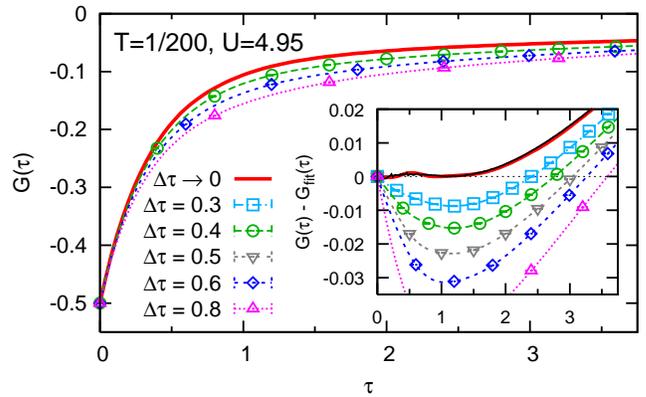}
    \caption{(Color online) Green function $G(\tau)$ for $T=1/200$,
      $U=4.95$: HF-QMC data of Fig.\ \ref{fig:GtauB200}
      (symbols) with interpolating fits (dashed/dotted lines) and
      numerically exact result after extrapolation $\dt\to 0$ (thick
      solid line). Inset: Green functions relative to $G_{\text{fit}}$;
      thin solid line: CT-QMC result \cite{Werner06}.}\label{fig:GsplineB200}
  \end{figure}

\begin{figure}
    \includegraphics[width=\columnwidth]{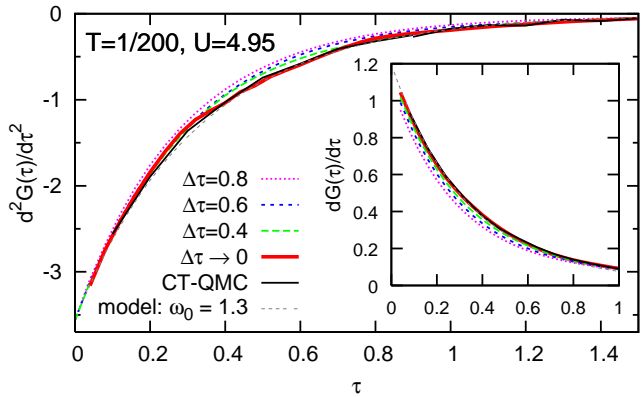}
    \caption{(Color online) Second and first order
   derivatives (main panel/\,inset) of the Green functions of Fig.\
      \ref{fig:GsplineB200}.}\label{fig:GderivB200}
  \end{figure}

  In a second step, the averaged QMC data sets (for different $\dt$)
  need to be transformed to a common grid. For this purpose, optimized
  two-pole approximations (as specified above) are obtained for each
  data set. Each difference $G_{\text{QMC}}-G_{\text{model}}$ (defined
  only on the specific grid $\{l\dt\}_{l=0}^\Lambda$) is interpolated
  by a natural cubic spline which is evaluated on a fixed, much finer
  grid; finally, $G_{\text{model}}(\tau)$ is added to these
  intermediate results \cite{fn:high-freq}. As seen in Fig.\
  \ref{fig:GsplineB200},
  these interpolated curves (dashed and dotted lines in the main panel)
  are smooth and vary systematically, with a discretization dependence
  which vanishes at small $\tau$. 

  In the third step and final step, least squares fits are performed
  for $\log[-G(\tau)]$ at each value of $\tau$ (on the fine grid),
  taking quadratic and quartic contributions in $\dt$ into account
  \cite{fn:weighting}. The result of this procedure is shown as thick
  solid line in Fig.\ \ref{fig:GsplineB200}. In the inset, the
  approximate low-$\tau$ asymptotics of the Green function,
  \[
  G_{\text{fit}}(\tau)=-0.5\, \exp\big[-2.36\,\tau\, (1 + 0.36\,\tau + 0.16\,\tau^2)^{-1}\big]\,
  \]
  have been subtracted. At this scale, the discretization error in the
  QMC data is clearly visible; in contrast, our extrapolated result
  (thick solid line) is hardly distinguishable from a continuous-time
  QMC estimate \cite{Werner06} (thin solid line). The fluctuations (on
  a much larger $\tau$ scale for HF-QMC than for CT-QMC) suggest that
  both results have similar precision. The competitiveness of HF-QMC
  (plus extrapolation) at this extremely low temperature, i.e., for
  rather coarse $\dt$ grids, may appear surprising. However, as
  shown in the inset of Fig.\ \ref{fig:GsplineB200}, the strongest
  absolute deviations of the raw HF-QMC data occur at $\tau\approx 1$
  (while the relative errors are largest at $\tau\approx 2$), i.e., at
  a position independent of $\dt$.  This establishes that the rapid
  initial decay of $G(\tau)$ does not set the scale for useful values
  of $\dt$ in HF-QMC and explains the good performance of our extrapolation
  procedure.

%%%%%%%%%%%%%%%%%%%%%%%%%%%%%%%%%%%%%%%%%%
  The uniform convergence of the (interpolated) HF-QMC Green
  functions, in turn, may be traced back to the fact that their
  curvatures are asymptotically exact for $\tau\to 0$. Indeed, the
  second derivatives $d^2 G(\tau)/d\tau^2$ are visibly $\dt$ dependent
  in the main panel of Fig.\ \ref{fig:GderivB200}
  only for $\tau\approx 0.5$; they all approach the correct asymptotic
  limit $d^2 G(\tau)/d\tau^2|_{\tau=0+}=(1+U^2/4)/2$ and agree at
  large $\tau$ within statistical errors. In contrast, as shown in the
  inset of Fig.\ \ref{fig:GderivB200}, the Trotter errors of the first
  derivatives do not vanish at $\tau\to 0$; concurrently, the
  asymptotic exact value $d G(\tau)/d\tau|_{\tau=0+}$ is nonuniversal,
  i.e., dependent on temperature and phase (metallic or insulating).
  Note that the extrapolated HF-QMC results in Fig.\
  \ref{fig:GderivB200} (thick solid lines) are hardly distinguishable
  from the corresponding CT-QMC results (thin solid lines).

  The specific form of the Green function extrapolation procedure
  detailed above is based on the insight that the Green function
  varies on a logarithmic scale. This is clearly seen in Fig.\
  \ref{fig:GtauB45U5}
  for a moderately low temperature $T=1/45$ and similarly strong
  interaction $U=5$ (close to the thermodynamic Mott transition):
  While all curves become indistinguishable for $\tau\to 0$, both
  differences between metal and insulator and the $\dt$ dependence in
  the insulating phase involve many orders of magnitude; in the latter
  case, even the error bars span nearly an order of magnitude.
  Obviously, a direct extrapolation $\dt\to 0$ of the insulating Green
  function (instead of $\log[-G(\tau)]$) would be hopeless for
  $\tau\gtrsim 5$. Note that our choice guarantees the correct sign
  for the extrapolated Green function. 

\begin{figure}
    \includegraphics[width=\columnwidth]{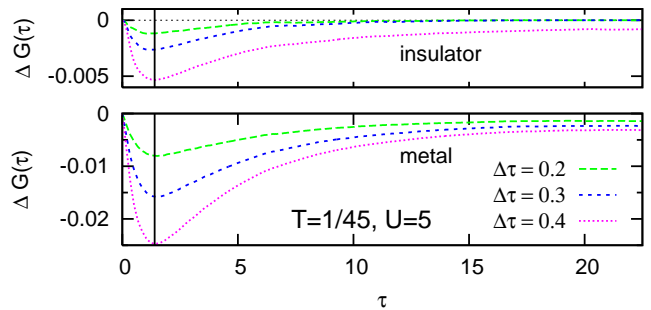}
    \caption{(Color online) Green functions: absolute deviations of
      finite-$\dt$ data of Fig.\ \ref{fig:GtauB45U5} from the
      extrapolated results for insulating/metallic phase (upper/lower
      panel).}\label{fig:GdiffB45U5}
  \end{figure}

  As seen in Fig.\ \ref{fig:GdiffB45U5}, 
  the absolute Trotter errors (i.e., the deviations at finite $\dt$
  from the numerically exact results) of all data sets peak, again, at
  the same position (here $\tau\approx 1.2$, indicated by vertical
  lines). The remarkable constancy of this peak position, even across
  phase transitions, implies that the extrapolation technique can be
  trusted to capture the low-$\tau$ features of the Green function
  without bias, even for relatively coarse grids $\dt$. In comparison,
  the region $\tau\approx \beta/2$ is more challenging (in the
  insulating phase): since HF-QMC results for $\tau\gtrsim 0.4$
  (dotted line in Fig.\ \ref{fig:GtauB45U5}) are too metallic, a good
  resolution of the exponential decay of $G(\tau)$ (see lower solid line
  in Fig.\ \ref{fig:GtauB45U5}) can only be expected from
  extrapolations which include small discretizations $\dt\lesssim
  0.3$.

%%%%%%%%%%%%%%%%%%%%%%%%%%%%%%%%%%%%%%%%%%%%%%%%%%%%%%%%%%%%
%%%%%%%%%%%%%%%%%%%%%%%%%%%%%%%%%%%%%%%%%%%%%%%%%%%%%%%%%%%%

\begin{figure}
    \includegraphics[width=\columnwidth]{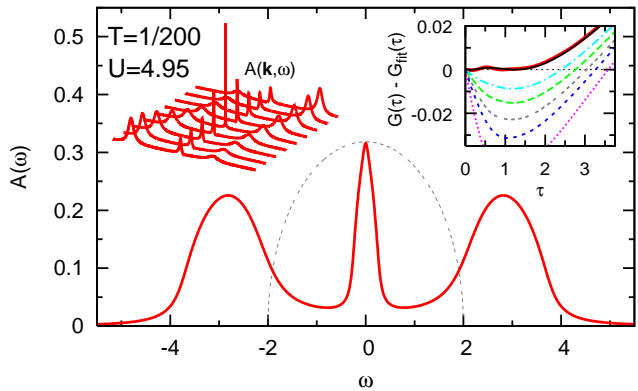}
    \caption{(Color online) Main panel: Local spectral function
      $A(\omega)$ for $T=1/200$, $U=4.95$ (solid line) in comparison
      with noninteracting spectrum (dashed line). Left inset:
      momentum-resolved spectra $A(\vk,\omega)$ in same energy range
      for momenta $\vk=\Gamma$ to $\vk=X$. Right inset: same data as
      in inset of Fig.\ \ref{fig:GsplineB200} except for the thin
      solid line which corresponds here to the spectrum of the main
      panel via Eq.\ \ref{eq:Gtau}.}\label{fig:dosB200}
  \end{figure}

  {\it Spectra on the real axis --} So far, we have only specified how
  to extract Green functions in the imaginary-time domain. The
  analytic continuation of this data to the real axis is still a
  highly nontrivial task; in fact, the extrapolated imaginary-time
  data arguably deserves even more careful and sophisticated
  continuation procedures than regular HF-QMC results, since -- for
  the first time -- the Green functions are exact within error bars.
  Adapted maximum entropy procedures and methods for efficient error
  analysis will be discussed elsewhere \cite{BluemerPrep}. In the case
  of very precise data (as in our examples), however, one may also
  neglect the small statistical errors and use Pad\'e methods. 
  Specifically, the local spectral function
  shown in the main panel of Fig.\ \ref{fig:dosB200}
  has been obtained via a Pad\'e approximant \cite{Vidberg77} to the
  imaginary-frequency self-energy $\Sigma(i\omega_n)$ using the first
  199 positive Matsubara frequencies. This procedure is stable [as has
  been checked via sum rules for $\Sigma(\omega)$ and $G(\omega)$ and
  by varying the Pad\'e parameters] and asymptotically exact in the
  noninteracting limit. In addition, it gives direct access also to
  momentum-resolved spectra (corresponding to ARPES measurements), as
  visualized in the left inset of Fig.\ \ref{fig:dosB200}.

  The expected pinning of the quasiparticle peak height in $A(\omega)$
  to its noninteracting value (cf.\ dashed line in main panel) is
  accurately observed; the spectrum is also reasonably smooth without
  any indications of artefacts. $A(\vk,\omega)$ clearly resolves the
  dispersions of the heavy quasiparticles (with $m^*/m\approx 9$) and
  of the incoherent Hubbard bands. So the results appear reasonable --
  but are they significantly better that those attainable using
  conventional HF-QMC/MEM procedures?  This is indeed the case, as
  demonstrated in the right hand inset of Fig.\ \ref{fig:dosB200}:
  here the thin solid line, computed from the final spectrum via Eq.\
  \ref{eq:Gtau}, is hardly distinguishable from the numerically exact
  $G(\tau)$ (thick line; cf.\ Fig.\ \ref{fig:GsplineB200}) while the
  deviations of finite-$\dt$ data (on which spectra would be based in
  conventional methods) are larger by several orders of magnitude.
  Thus, the spectrum of Fig.\ \ref{fig:dosB200} is seen to be unbiased
  and numerically exact -- for a temperature which had been deemed out
  of reach of HF-QMC.

%%%%%%%%%%%%%%%%%%%%%%%%%%%%%%%%%%%%%%%%%%%%%%%%%%%%%%%%%%%%

  {\it Discussion --} We have presented a method that, finally, allows
  to obtain numerically exact Green functions and ($\vk$ resolved)
  spectral functions from regular Hirsch-Fye QMC calculations. Due to
  the simplifications in DMFT, transport properties such as the
  optical conductivity $\sigma(\omega)$ could be computed without
  additional effort. Since the method has been successfully tested
  also for doped systems \cite{BluemerPrep}, it should greatly improve
  the attainable quality of spectra, in particular in the context of
  LDA+DMFT calculations.  In principle, the recently developed
  continuous-time QMC methods should yield spectra of similar quality
  with comparable effort; however, this might require regularization
  procedures which fail in insulating phases \cite{Krivenko06}.
  Numerical renormalization group methods appear hampered by their
  coarse high-frequency resolution and systematic errors. It contrast,
  detailed comparisons of numerically exact HF-QMC spectra with ground
  state estimates from dynamical density-matrix renormalization group
  are expected to lead to new physical insight.  Specificly, one may
  hope to thereby resolve the still controversial question of how the
  structure of the Hubbard bands changes across the Mott transition.

%  {\it Acknowledgments --} 
Stimulating discussions with P.G.J.\ van
  Dongen and support by the DFG within the Collaborative Research Centre 
  SFB/TR 49 are gratefully acknowledged.


\begin{thebibliography}{16}
\bibitem{fn:high-freq} A similar procedure is used for Fourier transformations within our DMFT-QMC code \protect\cite{BluemerKnecht,Bluemeretc,Bluemer07}.
\bibitem{fn:weighting} For stabilization of the least-squares fits, lower-$\dt$ data is overweighted by rescaling errors by $\dt/\dt_{\text{min}}$.
\end{thebibliography}
\end{document}